\documentclass[12pt]{iopart}

\newcommand{\bea}{\begin{eqnarray}}    
\newcommand{\eea}{\end{eqnarray}}      
\newcommand{\be}{\begin{equation}}
\newcommand{\ee}{\end{equation}}
\newcommand{\bef}{\begin{figure}}
\newcommand{\eef}{\end{figure}}

\def\spose#1{\hbox to 0pt{#1\hss}}
\def\ltapprox{\mathrel{\spose{\lower 3pt\hbox{$\mathchar"218$}}
\raise 2.0pt\hbox{$\mathchar"10C$}}}
\def\gtapprox{\mathrel{\spose{\lower 3pt\hbox{$\mathchar"218$}}
\raise 2.0pt\hbox{$\mathchar"10E$}}}
\def\inapprox{\mathrel{\spose{\lower 3pt\hbox{$\mathchar"218$}}
\raise 2.0pt\hbox{$\mathchar"232$}}}

\usepackage{epsfig} 

\begin{document}

\title{The complex universe: recent observations and theoretical
  challenges}

\author{Francesco Sylos Labini} \address{Enrico Fermi Center, Piazza
  del Viminale 1 00184 - Rome - Italy and Istituto dei Sistemi
  Complessi CNR, - Via dei Taurini 19, 00185 Rome, Italy}
\ead{Francesco.SylosLabini@roma1.infn.it}

\author{Luciano Pietronero}
\address{Istituto dei Sistemi Complessi
  CNR,  Via dei Taurini 19, 00185 Rome, Italy and  Physics
  Department, University of Rome ``Sapienza'', Piazzale Aldo Moro 2
  00185 Rome, Italy}
  \ead{Luciano.Pietronero@roma1.infn.it}

\begin{abstract}
The large scale distribution of galaxies in the universe displays a
complex pattern of clusters, super-clusters, filaments and voids with
sizes limited only by the boundaries of the available samples. A
quantitative statistical characterization of these structures shows
that galaxy distribution is inhomogeneous in these samples, being
characterized by large-amplitude fluctuations of large spatial
extension. Over a large range of scales, both the average conditional
density and its variance show a nontrivial scaling behavior: at small
scales, $r<20$ Mpc/h, the average (conditional) density scales as
$r^{-1}$. At larger scales, the density depends only weakly
(logarithmically) on the system size and density fluctuations follow
the Gumbel distribution of extreme value statistics. These complex
behaviors are different from what is expected in a homogeneous
distribution with Gaussian fluctuations.  The observed density
inhomogeneities pose a fundamental challenge to the standard picture of
cosmology but it also represent an important opportunity which points
to new directions with respect to many cosmological puzzles. Indeed,
the fact that matter distribution is not uniform, in the limited range
of scales sampled by observations, rises the question of understanding
how inhomogeneities affect the large-scale dynamics of the
universe. We discuss several attempts which try to model
inhomogeneities in cosmology, considering their effects with respect
to the role and abundance of dark energy and dark matter.
\end{abstract}


\maketitle

\section{Introduction}

Cosmological observations are usually interpreted within a theoretical
framework based on the simplest conceivable class of solutions of
Einstein's law of gravitation.  Namely, by using the assumptions that
the universe is homogeneous and isotropic, one is able to work out,
from the Einstein's field equations, the dynamics of space-time. The
Friedmann-Robertson-Walker (FRW) geometry is derived under these two
assumptions and it describes the geometry of the universe in terms of
a single function, the scale factor, which obeys to the Friedmann
equations \cite{weinberg}. In this situation, the matter density is
constant in a spatial hyper-surface. On the top of the constant matter
density one can consider a statistically homogeneous and isotropic
small-amplitude fluctuations field. These fluctuations furnish the
seeds of gravitational clustering which will eventually give rise to
the structures we observe in the present universe. Evolution of
fluctuations is not considered to have a sensible effect on the
evolution of the space-time which is instead driven by the uniform
mean field.

While the simplicity of this scenario has its own appeal, in the
standard model of cosmology one has to conjecture the existence of two
fundamental constituents, if observational constraints are met, that
both have yet unknown origin: first, a {\it dominant repulsive}
component is thought to exist that can be modeled, for instance, by a
positive cosmological constant. The physical nature of this component,
named Dark Energy, is yet unknown and its abundance cannot be inferred
from a-priori principles, whilst it is widely believed that dark
energy is the biggest puzzle in standard cosmology today; e.g., the
value of the cosmological constant in cosmology seems absurdly small
in the context of quantum physics \cite{review_weinberg}.

 There is, secondly, a {\it non-baryonic} component that should
 considerably exceed the contribution by luminous and dark baryons and
 massive neutrinos. This component, named non-baryonic Dark Matter, is
 thought to be provided by exotic forms of matter, not yet detected in
 laboratory experiments.  The main peculiar property of this matter
 component is that of weakly interacting with radiation in order to
 met the observational constraints given by observations
 \cite{einastodarkmatter}.
According to the concordance model of standard cosmology, the
contribution of the former converges to about 3/4 and that for the
latter to about 1/4 of the total source of the standard cosmological
equations (Friedmann equations), while up to a few percent has to be
attributed to what is instead directly measurable by observations,
namely ordinary baryonic matter, radiation and neutrinos.

If the underlying cosmological model is not a perturbation of an exact
flat FRW solution, the conventional data analysis and their
interpretation is not necessarily valid and thus the estimations of
Dark Matter and Dark Energy can be questionable.  The breaking of the
FRW solution can be caused, for instance, by strong inhomogeneities of
large spatial extension in the matter distribution. If this were the
case, the theoretical problem would then concern of whether
inhomogeneous properties of the Universe can be described by the
strong FRW idealization and/or in which limit this would be so.

The question of whether observations of galaxy structures satisfy, on
some scales, the assumptions used to derive the FRW metric is thus
central. Surprisingly enough, studies of the large scale distribution
of matter in the universe, as sampled by galaxy structures, seem not
to be trivially compatible with such a theoretical scenario. Indeed,
more and more structures on larger and larger scales have been
discovered in the course of the last two decades with the advent of
the three dimensional maps of the large scale universe. There
structures were unexpected both because two-dimensional (angular)
surveys were rather uniform and because theoretical models were unable
to predict the existence of them. In many cases it was concluded that
the particular three-dimensional survey under consideration had picked
up a particularly ``rare'' fluctuation: this was in respect to the
Gaussian distribution of fluctuations predicted by theoretical models.
The statistical characterization of these structures, determining the
range of correlations and the amplitudes and sizes of inhomogeneities,
has thus posed a fundamental challenge to the standard picture of
cosmology.  The key-problem would be then to include these large
fluctuations in the cosmological dynamics in a coherent way.

As long as structures are limited to small sizes, and fluctuations
have low amplitude, one can just treat fluctuations as small-amplitude
perturbations to the leading order FRW approximation. However if
structures have ``large enough'' sizes and ``high enough'' amplitudes,
a perturbation approach may loose its validity and a more general
treatment of inhomogeneities needs to be developed. From the
theoretical point of view, it is then necessary to understand how to
treat inhomogeneities in the framework of General Relativity.  In this
context the first issue is whether inhomogeneities can be described by
the FRW idealization at least {\it on average}, by postulating that on
large enough scales uniformity is eventually reached.  In other words,
the key-question is: does an inhomogeneous model of the Universe at
relatively small scales and, uniform at large scales, evolves on
average like a homogeneous solution of Einstein's law of gravitation ?
Currently there is a wide discussion in the literature on this issue
because, in the framework of averaged cosmological equations that has
been provided by Buchert \cite{buchert}, it was found that a potential
way to explain Dark Energy (and possibly also Dark Matter) can be,
partially or fully, given by an effect of structure formation in an
inhomogeneous cosmology. Inhomogeneities may mimic the effect of Dark
Energy \cite{wiltshire}.

Thus, while observations of galaxy structures have given an impulse to
the search for more general solution of Einstein's equations than the
Friedmann one, it is now under an intense investigation whether such a
more general framework may provide a different explanation to the
various effects that, within the standard FRW model, have been {\it
  interpreted} as Dark Energy and Dark Matter.

In these proceedings we first review in Sect.\ref{galaxies} the
situation with respect to galaxy structures: their observations and
the analysis of their statistical properties. We then discuss in
Sect.\ref{assumptions} the implications of the results on galaxy
distribution with respect to the standard theoretical assumptions of
the FRW model. This allows us to properly frame the problem of
inhomogeneities from the point of view of theoretical modeling.  We
then draw, in Sect.\ref{conclusions}, our main conclusions.

\section{Size and amplitude of galaxy structures} 
\label{galaxies}

In one of his seminal papers, Gerard de Vaucouleurs \cite{devac1970}
put into an historical perspective the problem of galaxy large scale
structures and the question about the scale where galaxy distribution
turns to uniformity (where by uniformity it is meant the absence of
structures of large amplitude or, in other words, of the absence voids
--- see below).  He pointed out that observations have firstly found
that galaxies are not randomly distributed. Then, that in the fifties
the same property was assigned to cluster centers. Finally that at the
end of the sixties the discovery of super-clusters has still enlarged
the scale of structures in the universe thus pushing to larger and
larger scales the scale where the approach to uniformity occurs.  In
the last twenty years many observations have been dedicated to the
study of the three-dimensional\footnote{This is achieved by measuring
  the redshift $z$ of a galaxy in addition to its angular
  coordinates. The Hubble's law linearly relate a galaxy redshift to
  its distance $R=c/H_0 z$ where $c$ is the light speed and $H_0$, the
  Hubble constant, is an observationally determined parameter.}
large-scale distributions of galaxies \cite{cfa1,cfa2,pp,ssrs2,lcrs}.
In particular during the last decade two ambitious observational
programs have measured the redshift of more than one million objects
\cite{colless01,york}.  All these surveys have detected larger and
larger structures, thus finding that galaxies are organized in a
complex network of clusters, super-clusters, filaments and voids.

For instance the famous ``slice of the Universe'', that represented
the first set of observations done for the CfA Redshift Survey in 1985
\cite{dlhg85}, mapped spectroscopic observations of about 1100
galaxies in a strip on the sky 6 degrees wide and about 130 degrees
long. This initial map was quite surprising, showing that the
distribution of galaxies in space was anything but random, with
galaxies actually appearing to be distributed on surfaces, almost
bubble like, surrounding large empty regions, or ``voids.''.  The
structure running all the way across the survey between 50 and 100
Mpc/h\footnote{We use $H_0=100 h$ km/sec/Mpc for the value of the
  Hubble constant; $h$ is a parameter constrained by observations to
  be in range $[0.5,075]$. Note that 1 Mpc $\approx 3 \cdot 10^{24}$
  cm and the size of the universe is thought to be $\sim 4000$ Mpc/h}
was called the ``Great Wall'' and at the time of the discovery was the
largest single structure detected in any redshift survey. Its
dimensions, limited only by the sample size, are about $200\times 80
\times 10 $ Mpc/h, a sort of like a giant quilt of galaxies across the
sky \cite{gh89}. More and more galaxy large scale structures were
identified in the other redshift surveys such as the Perseus-Pisces
super-cluster \cite{pp} which is one of two dominant concentrations of
galaxies in the nearby universe. This long chain of galaxies lies next
to the the so-called Taurus void, which is a large circular void
bounded by walls of galaxies on either side of it. The void has a
diameter of about 30 Mpc/h.  Few years ago, in the larger sample
provided by the Sloan Digital Sky Survey (SDSS) \cite{york}, it has
been discovered the Sloan Great Wall \cite{sloangreatwall}, which is a
giant wall of galaxies and which is the largest known structure in the
Universe, being nearly three times longer than the Great Wall.

\subsection{A characteristic scale for galaxy clustering ?}

Despite the fact that large scale galaxy structures, of size of the
order of several hundreds of Mpc/h, have been observed to be the
typical feature of the distribution of visible matter in the local
universe, the statistical analysis measuring their properties has
identified a characteristic scale which has only slightly changed
since its discovery fourthy years ago in angular catalogs.
Surprisingly enough in these samples, where only the angular
coordinates of galaxies were measured, it was not evident at all the
complex network of structures subsequently discovered with the advent
of redshift surveys.

Indeed, the characteristic scale $r_0$, defined to be the one at which
fluctuations in the galaxy density field are about twice the value of
the sample density, was measured to be $r_0 \approx$ 5 Mpc/h in the
Shane and Wirtanen angular catalog \cite{tk69}. More recent
measurements of this scale
\cite{dp83,davis88,park,benoist,norbergxi01,
  norbergxi02,zehavietal02,zehavietal04} in the three dimensional
catalogs found that $r_0$ fluctuates in the range $r_0 \approx 5-15$
Mpc/h 
\footnote{However, depending on how the sample density is estimated,
  also much larger values can be obtained \cite{sdss_aea,sdss_epl}}.
This variation was then ascribed to a luminosity dependent effect ---
see e.g.  \cite{davis88,park,benoist,zehavietal02}.  The small value
of the scale $r_0$ seems not to characterize the spatial extension of
structures, which can be even two orders of magnitude larger.  Indeed,
$r_0$ is a scale related to a specific value of the amplitude of
density fluctuations relative to the average density and not to their
spatial extension.

To simply understand this situation we recall some elementary concepts
\cite{book}.  For stationary density fields the quantity $\langle
\rho(\vec{r}_1)\rho(\vec{r}_2) \rangle$ is called the complete 2-point
correlation function (where $\langle ... \rangle$ is to mean the
ensemble average). If the field is statistically homogeneous and
isotropic then this function depends on the scalar distance between
points $r_{12} = |\vec{r}_1 - \vec{r}_2|$. For a spatially uniform
density field, for which the ensemble average density is $\rho_0>0$,
it is useful to consider the reduced correlation function
\be C_2(r_{12})
= \langle (\rho(\vec{r}_1)-\rho_0) (\rho(\vec{r}_2) -\rho_0) \rangle
\;.
\ee 
This is the main function used to study spatial correlations
between fluctuations from the average. The dimensionless two-point
correlation function usually considered in the analysis of 
galaxy distributions is defined as 
\be 
\label{eq:xi} 
\xi(r) = \frac{C_2(r_{12})}{\rho_0^2} 
\equiv \frac{\langle \rho(\vec{r}_1) \rho(\vec{r}_2)  \rangle}{\rho_0^2}  -1
\;.
\ee 
Note that this well-defined only when $\rho_0>0$. This function is
simply related to the normalized mass variance in a volume $V(R)$ of
linear size $R$
\be 
\label{eq:sigma2} 
\sigma^2(R) = \frac{\langle M(R)^2\rangle - \langle M(R)
  \rangle^2} {\langle M(R) \rangle^2} \ee 
by the following relation \cite{book} 
\be 
\sigma^2(R) = \frac{1}{V^2(R)} \int_{V(R)}d^3r_1\int_{V(R)} d^3r_2
\xi(r_{12}) \;. 
\ee 
For a spatially uniform density field, the scale $r_*$ at which
fluctuations are of the order of mean, i.e.
\be
\sigma(r_*) =1 \;, 
\ee 
is proportional to the scale $r_0$ at which $\xi(r_0)=1$.

Consider now the case in which correlations have a finite range so
that
\be
\label{eq:exp}
\xi(r)=A \exp(-r/r_c) 
\ee 
where $r_c$ is the {\it correlation length}
of the distribution and $A$ is a constant. Structures of fluctuations have a size determined by $r_c$. This
length scale does not depend on the amplitude $A$ of the correlation
function but only on its rate of decay.  On the other hand the scale 
at which $\xi(r_0)=1$ is 
\be
r_0 = r_c \cdot \log(A) \;,
\ee  
and it depends on the amplitude $A$ of the correlation function.  Thus
the two scales $r_*$ and $r_c$ have a completely distinct meaning: the
former marks the crossover from large to small fluctuation while the
latter quantifies the typical size of clusters of {\it small
  amplitude} fluctuations. When $\xi(r)$ is a power-law function of
separation \footnote{Note that we are discussing the case in which the
  distribution is spatially uniform. This means that small amplitude
  fluctuations have power-law correlations. On the other hand, the
  estimator of the function $\xi(r)$ can display power-law correlation
  even when the distribution itself presents scaling behavior as in a
  fractal \cite{book}. In such a situation, however, the (conditional)
  density itself presents scaling behavior, and the $\xi(r)$ analysis
  does not provide a statistically meaningful information (see
  discussion below).}  (with an exponent in the range $0<\gamma<3$)
then the correlation length $r_c$ is infinite and there are clusters
of all sizes \cite{book}.

In a finite sample, to give a physical meaning to $r_0$ or to $r_*$
one needs to verify that the average density is a well-defined
concept. Indeed, if the ensemble density is asymptotically zero the
normalization of the amplitudes of correlations to this value is not
possible. Nevertheless, in such a situation, in a finite sample, one
estimates the average density to be positive, with its precise value
depending on the sample size \cite{book}. This estimation is however
biased by the finite size of the sample. Thus, prior to the analysis
of fluctuations with statistical quantities like those defined in
Eq.\ref{eq:xi} and Eq.\ref{eq:sigma2} it is necessary to carefully
investigate whether the average density is stable ``enough'' in the
available samples.  Note that there has been an intense debate in the
last decade concerning the {\it statistical methods} employed to
characterize galaxy structures
\cite{slmp98,rees,hogg,book,bt05,sdss_aea,sdss_epl,sdss_epl,2df_epl,2df_aea}.
This was originated when it was realized \cite{pie87,cp92} that the
normalization of two-point correlations to the sample average, as
usually done in the field, can be problematic.

We thus face two distinct questions: What is the typical size of
structures ?  What is the amplitude of fluctuations at a given scale ?
The extension of structures can be statistically measured by the range
of positive correlations. The characterization of the amplitude of
fluctuations can be instead achieved by study their (conditional) PDF
as we discuss below. The key-question we face concerns the
quantification of absolute fluctuations and not of relative ones
normalized to a sample estimation of the average density. the
distinction between these two cases becomes especially relevant when
the distribution is dominated by a few structures as in such a
situation the concept of average density may loose its statistical
meaning \cite{book}.

A simple way to perform this measurement consists in the comparison of
galaxy counts in different angular regions \footnote{For this test one
  may consider counts as a function of distance and or of apparent
  luminosity (magnitude): they both may quantify fluctuations in
  different sky regions \cite{book,gsl01}}.  Many authors found that
there are fluctuations of the order of $\sim 30\%$ on scales of the
order of $200$ Mpc in a number of different three dimensional and
angular catalogs
\cite{picard91,busswell03,frith03,frith06,kerscher98,chiaki} implying
that there is more excess large-scale power than detected by the
standard correlation function analysis \cite{2df_epl,2df_aea}.  As we
discuss below, large scale structures and wide fluctuations at scales
of the order of 100 Mpc/h or more are at odds both the small value of
the characteristic length scale $r_0$ and with the predictions of the
concordance model of galaxy formation
\cite{busswell03,frith03,frith06}.

 To clarify this puzzling situation, i.e. the coexistence of the small
 typical length scales measured by the two-point correlation function
 analysis with the large fluctuations in the galaxy density field on
 large scales as measured by the simple galaxy counts, one needs to
 consider in detail the assumptions and the limits of a statistical
 analysis by which the quantitative characterization of structures is
 performed. Before entering in such a discussion let us briefly review
 the main predictions of structure formation models in the standard
 cosmological scenarios.

\subsection{Predictions from structure formation models} 

In structure formation models gravitational clustering drives
non-linear structures formation from an initially uniform density
field.  Due to the small initial velocity dispersion, structure
formation occurs in a bottom-up manner and thus fluctuations remain of
small amplitude at large enough scales while they acquire, as time
evolves, a large relative amplitude on some small scales. In this
situation, given that the large-scale uniformity is preserved, the
average density is a well defined concept at all times and the
length-scale $r_0$ does identify the typical size of non-linear
structures. This scale represents one of the main predictions of
theoretical models which must be confronted with observations.

Theoretical models of galaxy formation, like the Cold Dark Matter
(CDM) one --- see e.g. \cite{peacock} ---, are able to predict the
scale $r_0$ once it is given the amplitude and correlation properties
of the density field fluctuations in the early universe.  Note that in
a CDM model the main dynamical role is played by non-baryonic dark
matter and the addition of dark-energy modifies mainly the global
dynamical properties of the cosmology. In other words, as mentioned in
the introduction, a CDM universe with a cosmological constant (i.e., a
$\Lambda$CDM model) is made of about 1/4 of non-baryonic dark matter
and 3/4 of dark energy, with a small fraction of ordinary baryonic
matter. The theoretical motivation for these dark substances, as
mentioned in the introduction, finds its roots in the needs for an
exotic matter which weakly interacts with radiation and in the
repulsive substance required to satisfy by constraints imposed by a
series of cosmological observations \cite{spergel}.

Indeed, in these models, density fluctuations in the early universe are
coupled with radiation. Therefore one may obtain the normalization of
the initial conditions by measuring the amplitude and correlation
properties of the anisotropies of the Cosmic Microwave Background
Radiation (CMBR). Then by calculating the evolution of small density
fluctuations in the linear perturbation analysis of a self-gravitating
fluid in an expanding universe, it is possible to predict the scale
$r_0$ today. This turns out, in current models as the $\Lambda$CDM
ones, to be $r_0 \approx 5$ Mpc/h \cite{springel05}.  On scales
$r<r_0$ models are unable to make precise predictions on the shape of
the correlation function because gravitational clustering in the
non-linear regime is difficult to be treated. Gravitational N-body
simulations are then used to investigate structure formation in the
non-linear phase.

In addition models predict that, for $r>r_0$, a precise type of small
amplitude fluctuations. It is thus possible to simply relate, for
$r>r_0$, by using the linear perturbation analysis mentioned above,
the properties of fluctuations in the present matter density field to
those in the early universe. CDM models predict that for $r_0 < r
<r_c$, fluctuations have very small amplitude and weak positive
correlations. The situation in this range of scales is
well-approximated by Eq.\ref{eq:exp} (see \cite{cdm_theo} for
details). The length-scale $r_c$ thus represents the cut-off in the
size of weak amplitude (positively correlated) structures in standard
models. In addition, for $r>r_c$ \footnote{The scale $r_c$ estimated
  from CMBR measurements to be $r_c \approx 100$ Mpc/h
  \cite{cdm_theo}.} models predict that the matter density field
presents a specific type of {\it anti-correlations} \cite{glass}. In
particular, in these models correlations and anti-correlations are
finely balanced in such a way that
\be
\int_0^{\infty} d^3r \xi(r) = 0 \;, 
\ee
which is a global constraint on the behavior of the two-point
correlation function $\xi(r)$ corresponding to the super-homogeneous
properties of cosmological density fields. In brief, this is global
condition on the correlation properties of the matter density field,
which can be understood as a consistency constraint in the framework
of FRW cosmology, and it corresponds to a very fine tuned balance
between negative and positive correlations of density fluctuations and
to the fastest possible decay of the normalized mass variance on large
scales (see \cite{glass,book,lebo} for a discussion on this topic).

The fundamental tests for current models of galaxy formation then
concern: (i) whether density fluctuations at large scales (i.e. $r>10$
Mpc/h) have small amplitude or not and (ii) whether there are
anti-correlations on scales $r > r_c \approx 100$ Mpc/h
\cite{cdm_theo}. The primary problem to be considered in this respect
concerns the statistical methods used to measure the amplitude of
fluctuations and the range of correlations.

\subsection{Spatial homogeneity and self-averaging properties}

The problem of the statistical characterization these structures in a
finite sample, of volume $V$ containing $M$ galaxies, can be rephrased
as the problem of measuring volume averaged statistical
quantities. The basic issue concerns whether these are meaningful
descriptors, i.e. whether they give or not stable statistical
estimations of ensemble averaged quantities \cite{sdss_aea}.  This
problem is particularly important when only a few large scale
structures are present in a sample, a situation that occurs when
correlations are long-ranged.

In general it is assumed that galaxy distribution is an ergodic
stationary stochastic process \cite{book}, which means that it is
statistically translationally and rotationally invariant, thus
avoiding special points or directions.  Stationary stochastic
distributions satisfy these conditions also when they have zero
asymptotic average density in the infinite volume limit \cite{book}.
The assumption of ergodicity implies that in a single realization of
the microscopic number density field $n(\vec{r})$ the average density
$n_0$ in the infinite volume is well defined and equal to the ensemble
average density \cite{book}.  The constant $n_0$ is strictly positive
for homogeneous distributions and can be asymptotically zero for
infinite inhomogeneous ones \cite{book}. The infinite volume limit
must be considered in the definition of probabilistic properties, but
in any real samples, one is concerned only with finite volumes and
statistical determinations.

In inhomogeneous systems, like fractals, unconditional quantities are
not well defined as these distributions are characterized by having a
(conditional) average density which scales with the sample size and
tends to zero as a power-law \cite{book}. In this situation only
conditional quantities can be well-defined from a statistical point of
view. In addition, by studying conditional properties one is able to
make reliable tests to determine whether a distribution is 
spatially uniform. 

The simplest one is the conditional density \cite{book}: its
ensemble average can be defined as
\be
\langle n(r) \rangle_p = \frac{\langle n(r)n(0) \rangle}
{\langle{n}\rangle} \;.
\ee
It measures the density of points at distance $r$ from a point of the
distribution. For a fractal object one has $\langle n(r) \rangle_p
\propto r^{D-3}$, where $0<D<3$, so that it tends to zero in the
infinite volume limit making the definition of the correlation
function in Eq.\ref{eq:xi} meaningless.  The statistical estimator of
this latter statistics, in a finite sample, can be written as
\be \overline{\xi(r)} = \frac{
  \langle n(r) \rangle_p } {n_S} -1 
\ee 
where $n_S = \langle n(R_s) \rangle_p$ is the estimation of the
average density at the scale $R_s$ of the sample itself. It is clear
when $\langle n(r) \rangle_p$ has a power-law behavior,
$\overline{\xi(r)}$ is dependent on the sample size, resulting in an
intrinsic bias of this function \cite{book}.

Given that the finite sample estimation of the average mass density is
positive (unless the sample is empty), for inhomogeneous distributions
the relative error with respect to the ensemble value (i.e., zero) can
be arbitrarily large \cite{book}.  This situation occurs as long as
the sample size is smaller than the scale $\lambda_0$ at which the
distribution eventually turns to homogeneity, i.e.  beyond which
density fluctuations are small and the conditional density becomes
flat
\[
\langle n(r) \rangle_p \approx  const. \;,
\]
so that $D=3$ for $r>\lambda_0$ \cite{book}.  In the finite sample
analysis it is then necessary to study the conditional scaling
properties of statistical quantities, by an analysis of fluctuations
and correlations which explicitly considers whether a distribution can
be or not spatially homogeneous.

There is however an additional problem to be carefully considered when
dealing with inhomogeneous distributions in finite samples. Indeed,
one has to test whether local sample fluctuations allow the
determination of average conditional quantities
\cite{sdss_epl,sdss_aea,copernican}.  Indeed, statistical properties
are determined by making averages over the whole sample
volume~\cite{book}.  In doing so one implicitly assumes that a certain
quantity measured in different regions of the sample is statistically
stable, i.e., that fluctuations in different sub-regions are described
by the same PDF.  However it may happen that measurements in different
sub-regions show systematic (i.e., not statistical) differences, which
depend, for instance, on the spatial position of the specific
sub-regions.  In this case the considered statistic is not
statistically stationary in space, the fluctuations systematically
differ in different sub-regions and whole-sample average values are
not meaningful descriptors \cite{book,sdss_aea}. For a stationary
stochastic point process this situations corresponds to the lack of
self-averaging properties \cite{sdss_aea}.  On the other hand a
systematic dependence of the PDF on the specific position of the
sample volume may correspond to the lack of stationarity (i.e., lack
of statistical translational invariance) of the distribution, a
situation that occurs, for instance, when there is a center breaking
overall translational invariance \cite{copernican}.

In order to define a quantitative test for self-averaging, let us
remind that a crucial assumption usually used is that stochastic
fields are required to satisfy spatial {\em ergodicity}.  Let us take
a generic observable \[ {\cal F}= {\cal
  F}(\rho(\vec{r}_1),\rho(\vec{r}_2),...)\] function of the mass
distribution $\rho(\vec{r})$ at different points in space
$\vec{r}_1,\vec{r}_2,...$.  Ergodicity implies that \be \left<{\cal
  F}\right> = \overline{{\cal F}} = \lim_{V \rightarrow \infty}
\overline{{\cal F }}_V \;, \ee where $\overline{{ \cal F}}_V$ is the
spatial average in a finite volume $V$ \cite{book}.
When considering a finite sample realization of a stochastic process,
and thus statistical estimators of asymptotic quantities, the first
question to be sorted out concerns whether a certain observable is
self-averaging in a given finite volume \cite{Aharony,sdss_aea}.  In
general a stochastic variable $\cal{F}$ is self-averaging if ${\cal F}
= \left< {\cal F}\right> $ (see \cite{sdss_aea} for a more detailed
discussion).
Thus if this is ergodic, $\overline{{\cal F}} = \left< {\cal F}\right>$,
then it is also self-averaging as $\overline{{ \cal F}} = \left<
\overline{{ \cal F} } \right>$: 
finite sample spatial averages must be self-averaging in order to
satisfy spatial ergodicity.

A simple test to determine whether a {\it distribution} is stationary
and self-averaging in a given sample of linear size $L$ consists in
studying the probability density function (PDF) of conditional
quantities ${ \cal G}$ (which contains, in principle, all information
about moments of any order) in sub-samples of linear size $L' < L$
placed in different and non-overlapping spatial regions of the sample
(i.e., $S_1,S_2,...S_N$).  That the self-averaging property holds is
shown by the fact that $P({ \cal G},L';S_i)$ is the same, modulo
statistical fluctuations, in the different sub-samples, i.e., \be
\label{selfavtest}
 P({ \cal G},L';S_i) \approx P({ \cal G},L';S_j) \; \forall i \ne j \; .
\ee
%
On the other hand, if determinations of $P({ \cal G},L'; S_i)$ in
different sample regions $S_i$ show {\it systematic} differences, then
there are two different possibilities: (i) the lack of the property of
stationarity or (ii) the breaking of the property of self-averaging due
to a finite-size effect related to the presence of long-range
correlated fluctuations.  Therefore while the breaking of statistical
homogeneity and/or isotropy imply the lack of self-averaging property
the reverse is not true. However, if the determinations of the spatial
averages give sample-dependent results, this implies that those
statistical quantities do not represent the asymptotic properties of
the given distribution \cite{sdss_aea}.

As mentioned above, to test statistical and spatial homogeneity it is
necessary to employ statistical quantities that do not require the
assumption of spatial homogeneity inside the sample and thus avoid the
normalization of fluctuations to the estimation of the sample average
\cite{sdss_aea}.  We therefore consider the statistical properties of
the stochastic variable defined by number of points $N_i(r)$ contained
in a sphere of radius $r$ centered on the $i^{th}$ point.  This
depends on the scale $r$ and on the spatial position of the $i^{th}$
sphere's center, namely its radial distance $R_i$ from a given origin
and its angular coordinates $\vec{\alpha}_i$.  Integrating over
$\vec{\alpha}_i$ for fixed radial distance $R_i$, we obtain that
$N_i(r)=N(r; R_i)$ \cite{sdss_aea,sdss_epl}.

\subsection{Results in galaxy catalogs}

\begin{figure}
\begin{center}
\includegraphics*[angle=0, width=0.85\textwidth]{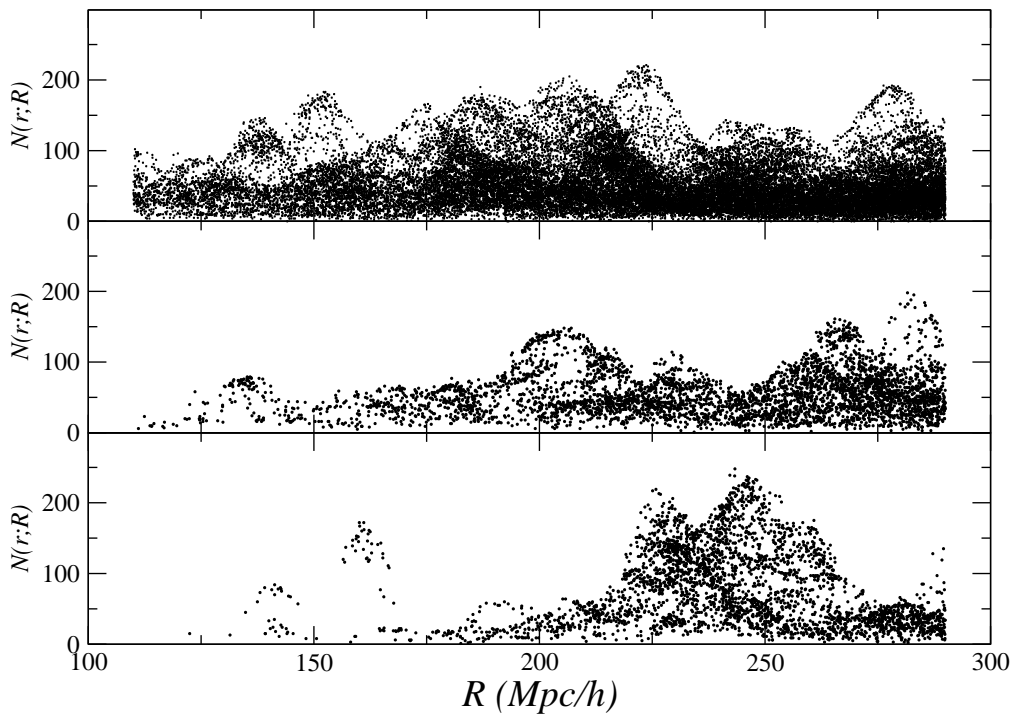} 
\caption{Behavior of $N(r;R_i)$ in a SDSS sample and in the
          three different regions for $r=10$ Mpc/h (R1 top, R2 Middle
          and R3 bottom). (From \cite{sdss_aea}).}
\label{SLVL2-R1R2R3}
\end{center}
\end{figure}

The analysis of $N_i(r;R)$ is found to be very efficient in mapping
large scale structures which manifest themselves as large fluctuations
in the $N_i(r;R)$ distributions for different positions $i$ and
spheres radii $r$.  For instance by studying this random variable in
various three-dimensional slices of the SDSS samples we identify a
giant filament covering, in the largest contiguous volume of the
survey, more than 400 Mpc/h at a distance $R\sim 500$ Mpc/h from us.
In different sub-samples this analysis reveals a variety of
structures, showing that large density fluctuations are quite
typical. An example is shown in Fig.\ref{SLVL2-R1R2R3} which displays
the behavior of $N(r;R_i)$ in three different regions of a sample
extracted from the SDSS.

\begin{figure}
\begin{center}
\includegraphics*[angle=0, width=0.85\textwidth]{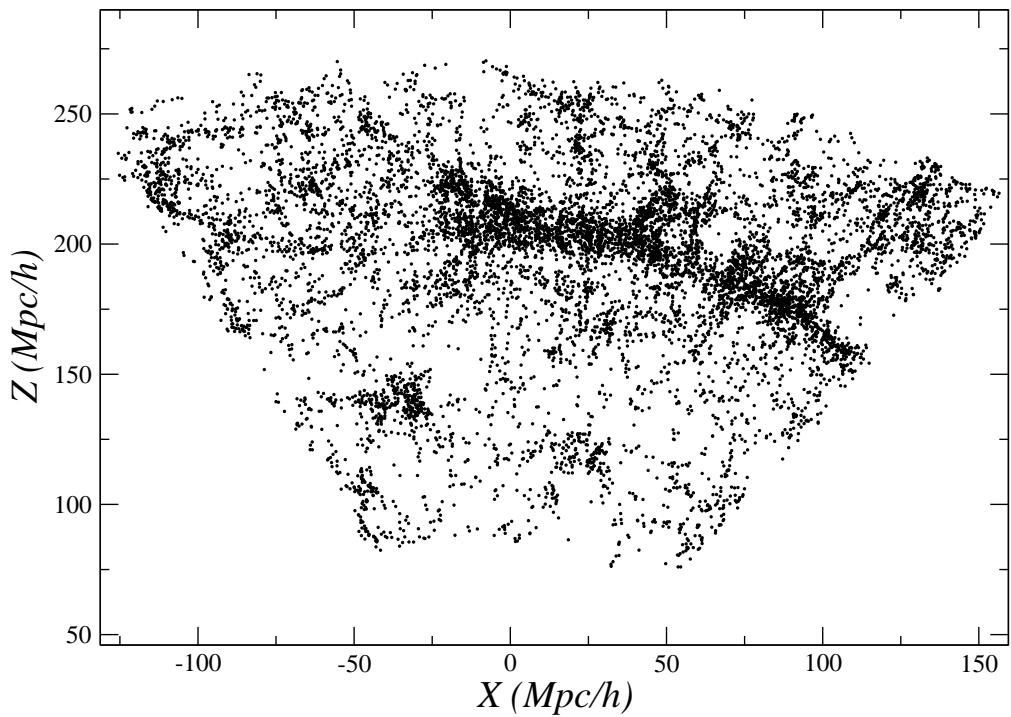}
        \caption{Projection on the $X-Z$ plane of a SDSS sample in
          which the Sloan Great Wall appears as a long filament of
          galaxies. (From \cite{sdss_aea}). }
\label{SDSSGw}
\end{center}
\end{figure}

One may note that this analysis is more powerful in tracing large
scale galaxy structures than the simple counting as a function of
radial distance. Indeed, one may precisely describe the sequence of
structures and voids characterizing the samples and, by changing the
sphere radius $r$, one may determine the situation at different
spatial resolutions.  For instance the distribution in a certain
region (see the bottom panel of Fig.\ref{SLVL2-R1R2R3}) is dominated
by a single large scale structure, which is known as the SDSS Great
Wall \cite{sloangreatwall}.  (In Fig.\ref{SDSSGw} we show the
projection on the $x-z$ plane of a sample where the SDSS Great Wall is
placed in the middle).  In some regions, which cover a small enough
sky area, one is a able to well isolate structures at different
distances, while the largest contiguous region, which covers a solid
angle about six times larger than the other two sky areas, the signal
is determined by the superposition of many structures of different
amplitude and at different scales.  By the simple visual comparison of
the profile in the different regions we can conclude that, although
the Great Wall is a particularly long filament of galaxies, it
represents a typical persistent fluctuation.

\begin{figure}
\begin{center}
\includegraphics*[angle=0, width=0.85\textwidth]{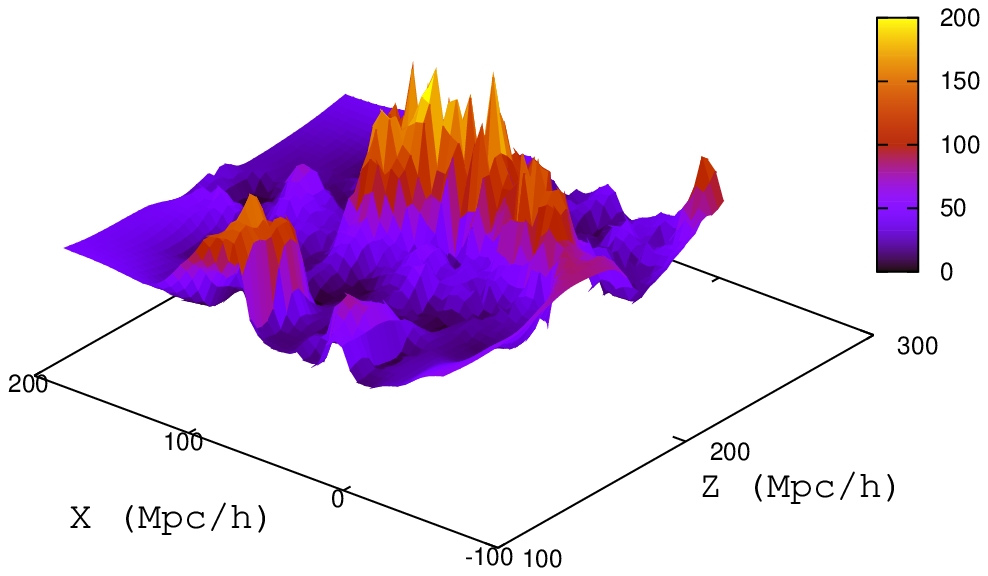}
\caption{Three dimensional representation of the SL analysis with
  $r=10$ Mpc/h for R3VL2. The $x,z$ coordinates of the sphere center
  define the bottom plane and on the vertical axis we display the
  intensity of the structures, the conditional number of galaxies
  $N_i(r)$ contained in the sphere of radius $r$.  The SDSS Great Wall
  is clearly visible as a coherent structure { of large amplitude},
  similar to a mountain chain, extending all over the sample. (From
 \cite{sdss_aea}).}
\label{SDSSGw_col}
\end{center}
\end{figure}

\begin{figure}
\begin{center}
\includegraphics*[angle=0, width=0.85\textwidth]{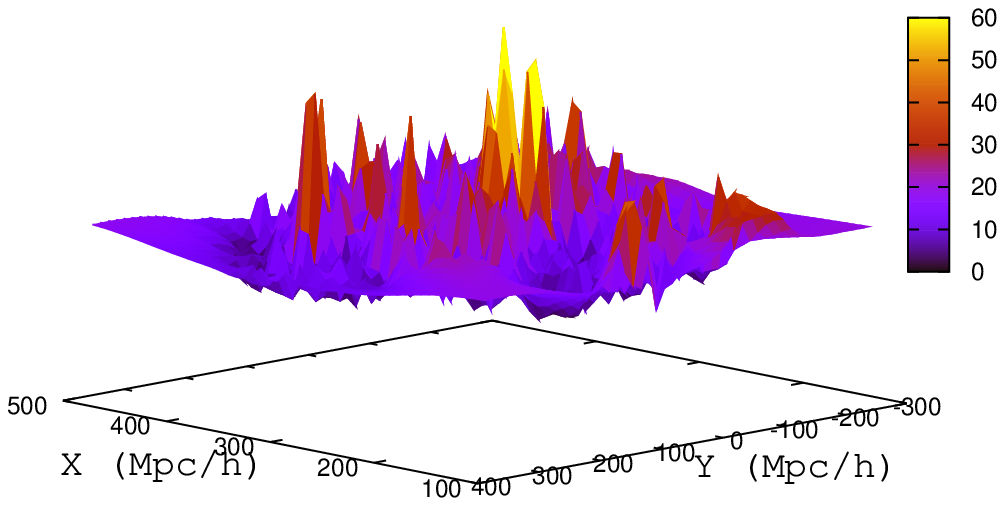}
\caption{Analysis similar to that shown in Fig.\ref{SDSSGw_col} but
  for a sample of the 2dFGRS.  On the $X$ and $Y$ axes the coordinate
  of the center of a sphere of radius $r=10$ Mpc/h (centered on a
  galaxy) is reported and on the $Z$ axis the number of galaxies
  inside it.  The mean thickness of this slice is about 50 Mpc/h.
  Large fluctuations in the density field are located in the
  correspondence of large-scale structures.  (From \cite{2df_aea}).}
\label{fig_sgp550_color_sl}
\end{center}
\end{figure}

More information about structures amplitude and location is provided
by the full $N_i(r) = N(r; x_i,y_i,z_i)$ data, where $(x_i,y_i,z_i)$
are the Cartesian coordinates of the $i^{th}$ center. In order to
illustrate this point, we have chosen in Fig.\ref{SDSSGw_col} a three
dimensional representation where on the bottom plane we use the $x,z$
Cartesian coordinates of the sphere center and on the vertical axis we
display the intensity of the structures, i.e.  the conditional number
of galaxies contained in the sphere of radius $r$.  (On the $y$
direction the thickness of the sample is small, i.e. $\Delta y \approx
15$ Mpc/h).  One may note that that the SDSS Great Wall is clearly
visible as a coherent structure similar to a mountain chain, extending
all over the sample.  It is worth noticing that profiles similar to
those shown in Fig.\ref{SLVL2-R1R2R3} and Fig.\ref{SDSSGw_col} have
been found also in the 2dFGRS \cite{2df_epl,2df_aea}
supporting the fact that these fluctuations are quite typical of
galaxy distribution (see Fig.\ref{fig_sgp550_color_sl}).

\begin{figure}
\begin{center}
\includegraphics*[angle=0, width=0.85\textwidth]{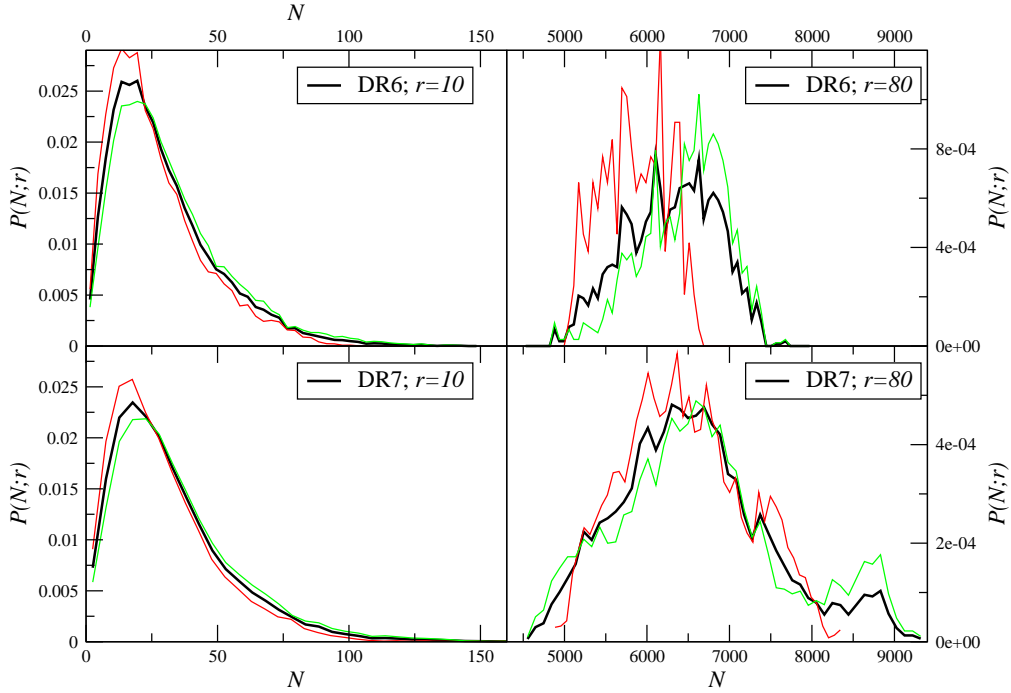}
        \caption{ PDF of the conditional galaxy counts in spheres, in
          the sample defined by $R \in [125, 400]$ Mpc/h and $M \in
          [−20.5,− 22.2]$ in the DR6 (upper panels) and DR7 (lower
          panels) data, for two different values of the sphere radii
          $r = 10$ Mpc/h and $r = 80 $ Mpc/h. In each panel, the black
          line represents the full-sample PDF, the red line (green)
          the PDF measured in the half of the sample closer to
          (farther from) the origin (From \cite{copernican}).}
\label{fig:ss1}
\end{center}
\end{figure}

We now pass to the determination of the PDF of the conditional galaxy
counts in spheres $P(N;r)$, at different resolution $r$, separately in
two independent regions of a given sample placed at different radial
distances.  In a first case (left panels of Fig.\ref{fig:ss1}), at
small scales ($r=10$ Mpc/h), the distribution is self-averaging both
in the earlier data release of the SDSS (DR6 sample \cite{dr6}, that
covers a solid angle $\Omega_{DR6} =0.94$ sr.)  than in the sample
extracted from the final data release (DR7 \cite{dr7} with
$\Omega_{DR7}=1.85$ sr. $\approx 2 \times \Omega_{DR6}$ sr).  Indeed,
the PDF is statistically the same in the two sub-samples considered.
Instead, for larger sphere radii i.e., $r=80$ Mpc/h, (right panels of
Fig.\ref{fig:ss1}) in the DR6 sample, the two PDF show clearly a
systematic difference. Not only the peaks do not coincide, but the
overall shape of the PDF is not smooth and different. On the other
hand, for the sample extracted from DR7, the two determinations of the
PDF are in very good agreement.  We conclude therefore that, in DR6
for $r=80$ Mpc/h there are large density fluctuations which are not
self-averaging because of the limited sample volume \cite{sdss_aea}.
They are instead self-averaging in DR7 because the volume is increased
by a factor two.

The lack of self-averaging properties at large scales in the DR6
sample is due to the presence of large scale galaxy structures which
correspond to density fluctuations of large amplitude and large
spatial extension, whose size is limited only by the sample
boundaries.  The appearance of self-averaging properties in the larger
DR7 sample volumes is the {\it unambiguous proof} that the lack of
them is induced by finite-size effects due to long-range correlated
fluctuations \cite{copernican}. The lack of self-averaging does not
allow one to characterize the nature of fluctuations; this is however
a clear indication that the distribution has not reached spatial
uniformity.

\begin{figure}
\begin{center}
\includegraphics*[angle=0, width=0.85\textwidth]{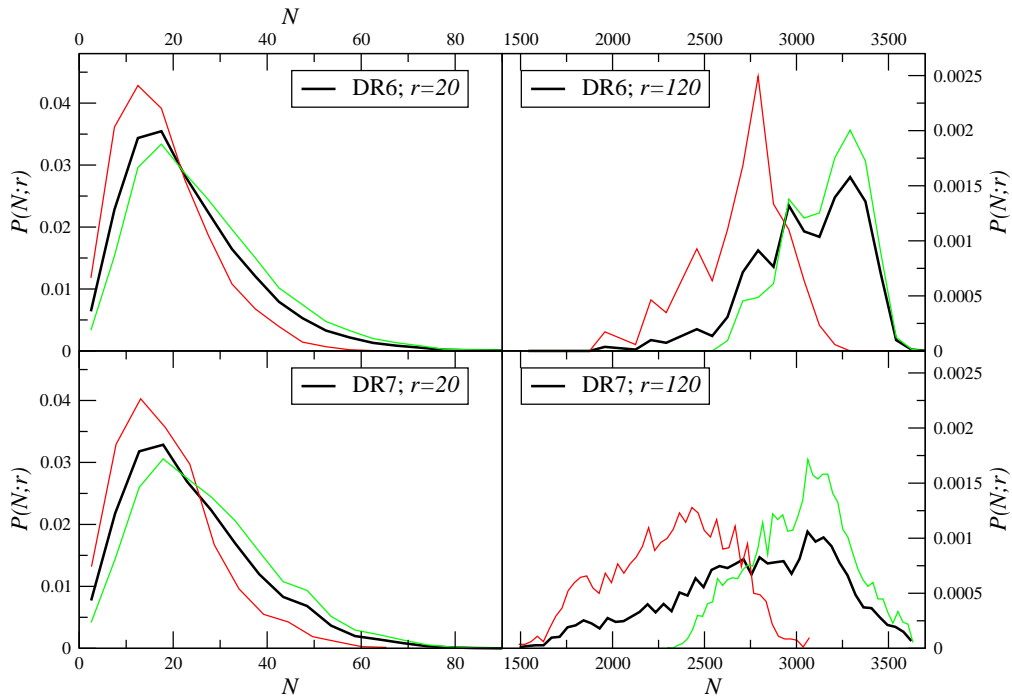} 
        \caption{The same of Fig.\ref{fig:ss1} but for the sample
          defined by $R \in [200, 600]$ Mpc/h and $M \in [−21.6,−
            22.8]$ and for $r = 20, 120$ Mpc/h (From
          \cite{copernican}).}
\label{fig:ss2}
\end{center}
\end{figure}

In the deepest sample we consider, which include mainly bright
galaxies, the breaking of self-averaging properties does not occur as
well for small $r$ but it is found for large $r$. This can be due to
the same effects i.e., that the sample volumes are still too small as
even in DR7 for $r =120$ Mpc/h we do not detect self-averaging
properties (right panels of Fig.\ref{fig:ss2}).  Other radial
distance-dependent selections, like galaxy evolution \cite{loveday},
could in principle result in an effect in the same direction.  Even in
such a situation, our conclusion would be unchanged, i.e. on large
enough scales self-averaging is broken. The reason why it is broken is
then different: instead of the effect intrinsic fluctuations in the
galaxy distribution, the result of a redshift-dependent effect.
Because of these large fluctuations in the galaxy density field,
self-averaging properties are well-defined only in a limited range of
scales. Only in that range it will be statistically meaningful to
measure whole-sample average quantities \cite{sdss_aea,gumbel}.

Let us now pass to the determination of the first moment of the
PDF, namely the conditional average density within radius
$r$  defined as
\begin{equation}
  \label{eq:condden}
  \overline{n(r)} = \frac{1}{V(r)} 
\frac{1}{M(r)} \sum_{i=1}^{M(r)} N_i(r) \;,
\end{equation}
where $\overline{n(r)}$ is, as discussed above, ``conditioned'' on the
presence of the central galaxy.  Then, the simplest quantity to
further characterize density fluctuations is the conditional variance,
or mean square deviation at scale $r$, i.e. the second moment of the
 PDF, which is defined as
\begin{equation}
  \label{eq:condvar}
  \sigma^2(r) \equiv \mathrm{var}\, [n(r)] = \frac{1}{M(r)}
  \sum_{i=1}^{M(r)} n_i^2(r) - {\overline{n(r)}}^2 \;.
\end{equation}

\begin{figure}
\begin{center}
\includegraphics*[angle=0, width=0.85\textwidth]{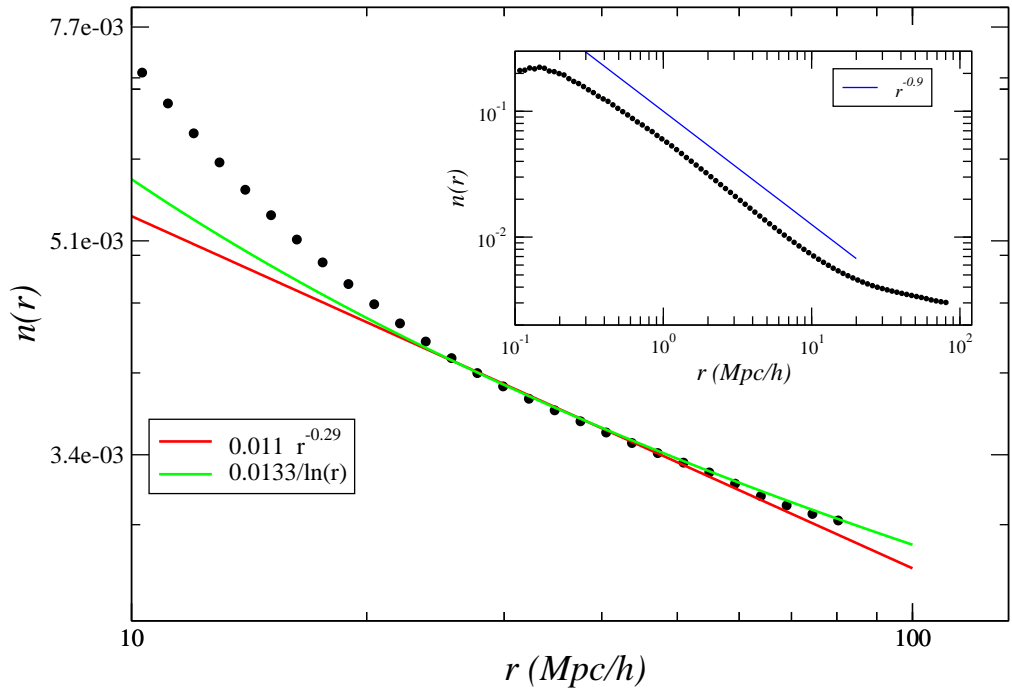}
\caption{Conditional average density $\overline n(r)$ of galaxies as a
  function of radius. In the inset panel the same is shown in the full
  range of scales. Note the change of slope at $\approx 20$ Mpc/h from
  $1/r$ to $1/r^{0.3}$ and also the lack of flattening up to $\approx
  80$ Mpc/h. (From \cite{gumbel}).}
\label{fig:condden}
\end{center}
\end{figure}

At small length scales ($r < 20$ Mpc/h) the conditional average
density shows a scaling behavior with an exponent close to minus one
(see Fig.\ref{fig:condden}).  This result is in agreement with the
ones obtained by the same method in a number of different samples (see
\cite{sdss_epl,sdss_aea,2df_epl,2df_aea} and references therein).
This scaling can be interpreted as a signature of fractality of the
galaxy distribution in this range of scale. In addition, this implies
that the distribution is not uniform at these scales, and thus the
standard two-point correlation function is substantially biased.


Then, the average conditional density (Eq.\ref{eq:condden}) at larger
scales ($r > 20$ Mpc/h ) shows a different $r$ dependence, as can be
seen in Fig.\ref{fig:condden}. Our best fit is
\begin{equation}
\label{densscale}
\overline{n(r)} \approx \frac{0.0133}{\log r} \;,
\end{equation}
that is the average density depends only weakly (logarithmically) on
$r$.  Alternatively, an almost indistinguishable power-law fit is
provided by
\begin{equation}
\label{densscale2}
\overline{n(r)} \approx 0.011 \times r^{-0.29} \;.
\end{equation}
We thus find a change of slope in the conditional average density in
terms of the radius $r$ at about $\approx 20$ Mpc/h. At this point the
decay of the density changes from an inverse linear decay to a slow
logarithmic one. Moreover, the density $\overline{n(r)}$ does not
saturate to a constant up to $\sim 80$ Mpc/h, {\it i.e.,} up to the
largest scales probed in this sample.  Note that up to $r=80$ Mpc/h
the number of points $M(r)$ is larger than $10^4$, so that the
statistics is sufficiently robust.

\begin{figure}
\begin{center}
\includegraphics*[angle=0, width=0.85\textwidth]{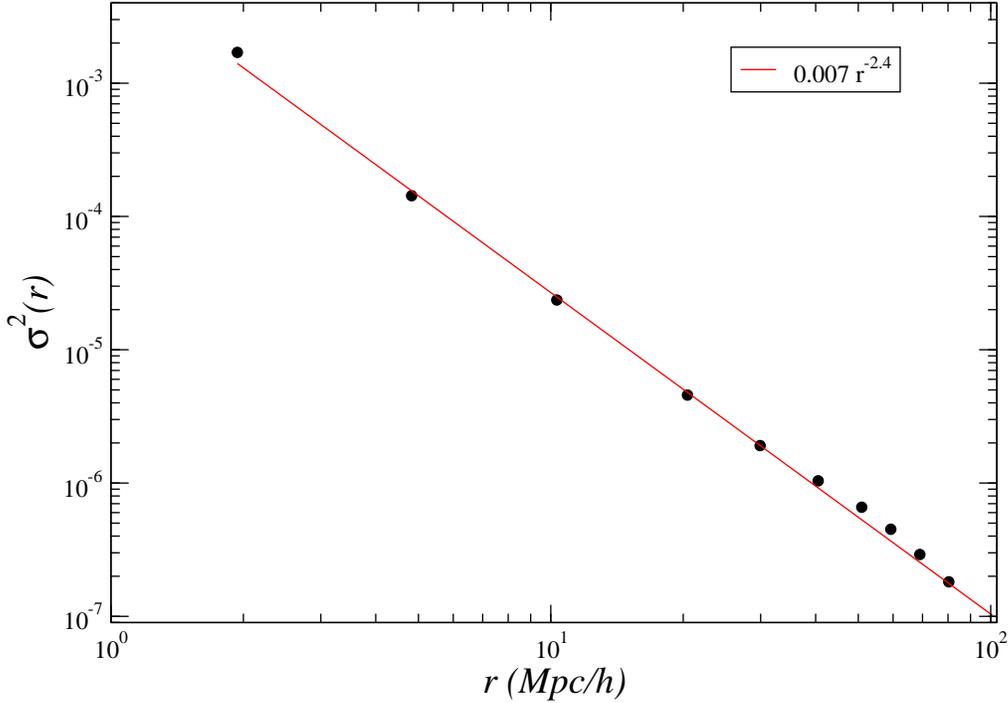}
\caption{Variance $\sigma^2$ of the conditional density $n_i(r)$ as a
  function of the radius.  Conversely, the corresponding variance of a
  Poisson point process would display a $1/r^3$ decay.  (From
  \cite{gumbel}).}
\label{fig:variance}
\end{center}
\end{figure}


This result is in agreement with a study of the SDSS-DR4 samples
\cite{dr4}, where, a similar change of slope was observed at about the
same scale $r \approx 20$ Mpc/h, together with quite large
fluctuations.  Indeed, some evidences were subsequently found to
support that the galaxy distribution is still characterized by rather
large fluctuations up to 100 Mpc/h, making it incompatible with
uniformity \cite{2df_epl,2df_aea,sdss_epl,sdss_aea,sdss_bao}.  In the
Luminous Red Galaxy (LRG) sample of SDSS, Hogg et al. \cite{hogg} also
found that the slope changes at $\sim$ 20 Mpc/h but then they claim to
detect a transition to uniformity at about 70 Mpc/h, We do not observe
in a clear way such a transition in the samples for which
self-averaging properties are satisfied.

Our best fit for the variance $\sigma^2(r)$ of the conditional density
(Eq.\ref{eq:condvar}) is (see Fig.\ref{fig:variance})
\begin{equation}
\label{denssigma2}
\sigma^2(r) \approx 0.007\times r^{-2.4} \;.
\end{equation}
Given the scaling behavior of the conditional density and variance, we
conclude that galaxy structures are characterized by non-trivial
correlations for scales up to $r \approx 80$ Mpc/h.

\begin{figure}
\begin{center}
\includegraphics*[angle=0, width=0.85\textwidth]{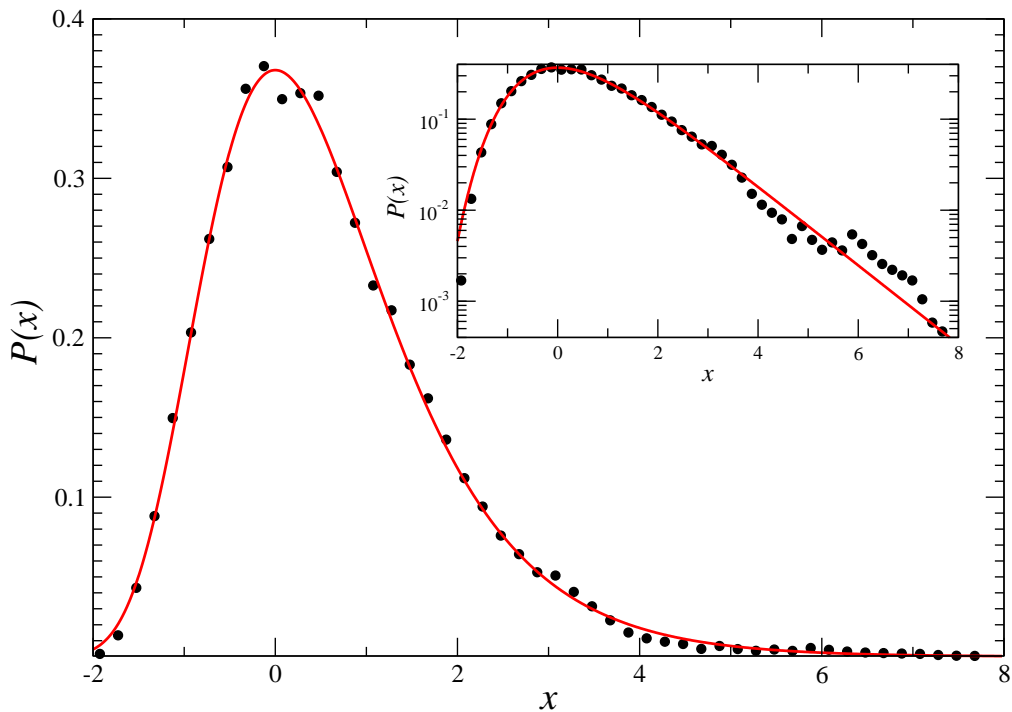}
\caption{One of the best fits is obtained for $r=20$. The data is
  rescaled by the fitted parameters $\alpha$ and $\beta$. The solid
  line corresponds to the parameter-less Gumbel distribution
  Eq.\ref{gumbelscaled}. The inset depicts the same on log-linear
  scale.(From \cite{gumbel}).}
\label{fig:best20}
\end{center}
\end{figure}

Let us now turn to the analysis of the PDF. It is well-known that away
from criticality, when correlations are long-ranged and the
correlation length diverges, any global (spatially averaged)
observable of a macroscopic system has Gaussian fluctuations, in
agreement with the central limit theorem (CLT). At criticality,
however, the correlation length tends to infinity, and the CLT no
longer applies. Indeed, fluctuations of global quantities in critical
systems usually have non-Gaussian fluctuations. The type of
fluctuations is characteristic to the universality class of the
system's critical behavior \cite{foltin94,racz02}. Generally when
correlations are long-ranged long-tailed distributions are found. In
this situation some moments of the distribution may diverge as there
is a finite probability to find fluctuations faraway from the most
probable value \cite{bouchoud}.

To fit experimental data, the (generalized) Gumbel PDF \cite{gumbel}
has often been used, where $a$ is a real parameter. For integer values
of $a$, this distribution corresponds to the $a$-th maximal value of a
random variable. The $a=1$ case corresponds to the Gumbel
distribution. Experimental examples for Gumbel or generalized Gumbel
distributions include power consumption of a turbulent flow
\cite{bramwell98}, roughness of voltage fluctuations in a resistor
(original Gumbel $a=1$ case) \cite{antal01}, plasma density
fluctuations in a tokamak \cite{milligen05}, orientation fluctuations
in a liquid crystal \cite{joubaud08}, and other systems cited in
\cite{bramwell09}.  The Gumbel distribution describing fluctuations of
a global observable was first obtained analytically in \cite{antal01}
for the roughness fluctuations of $1/f$ noise. Its relations to
extreme value statistics have been clarified \cite{bertin05,bertin06},
generalizations have appeared \cite{antal02}, and related finite size
corrections have been understood \cite{gyorgyi08}.

In a recent paper Bramwell \cite{bramwell09} conjectured that only
three types of distributions appear to describe fluctuations of global
observables at criticality. In particular, when the global observable
depends logarithmically on the system size, the corresponding
distribution should be a (generalized) Gumbel. For example the mean
roughness of $1/f$ signals depends on the logarithm of the observation
time (system size), and the corresponding PDF is indeed the Gumbel
distribution \cite{antal01}.

The Gumbel (also known as Fisher-Tippet-Gumbel) distribution is one of
the three extreme value distribution \cite{fisher28,gumbel58}. It
describes the distribution of the largest values of a random variable
from a density function with faster than algebraic (say exponential)
decay. When fluctuations are characterized by the Gumbel PDF the
situation is in between the Gaussian case, where all moments are
well-defined, and the case in which the tail of the PDF scales as a
power-law, i.e. the case in which several moments of the PDF may
diverge corresponding to an is extremely wild fluctuation field. It is
important to stress that models of galaxy formation predict
Gaussian-like fluctuations on sufficiently large scales. The key-issue
concerns indeed at which scales fluctuations show a Gaussian behavior,
which is a closely related problem to the scale at which the
distribution turns to spatial uniformity \cite{sdss_aea,sdss_epl}.

The Gumbel distribution's PDF is given by
\begin{equation}
\label{gumbel}
 P(y)= \frac{1}{\beta} 
\exp\left[ - \frac{y-\alpha}{\beta} - 
\exp\left( - \frac{y-\alpha}{\beta} \right) \right] \;.
\end{equation}
With the scaling variable
\begin{equation}
\label{scalevar}
 x = \frac{y-\alpha}{\beta}
\end{equation}
the density function (Eq.\ref{gumbel}) 
simplifies to the parameter-free Gumbel
\begin{equation}
\label{gumbelscaled}
 P(x) = e^{-x-e^{-x}}
\end{equation}
with (cumulative) distribution $e^{-e^{-x}}$.  Note that this
distribution corresponds to large extremes, while for low extreme
values, $x$ is used instead of $-x$ in the Gumbel distribution.

The mean and the standard deviation (variance) of the Gumbel
distribution (Eq.\ref{gumbel}) are respectively
\begin{equation}
\label{cumu}
 \mu = \alpha + \gamma \beta, \quad 
 \sigma^2 = (\beta\pi)^2/ 6 
\end{equation}
where $\gamma=0.85772\dots$ is the Euler constant. For the scaled
Gumbel (Eq.\ref{gumbelscaled}) the first two cumulants of Eq.\ref{cumu}
simplify to $\gamma$ and $\pi^2/6$.

\begin{figure}
\begin{center}
\includegraphics*[angle=0, width=0.85\textwidth]{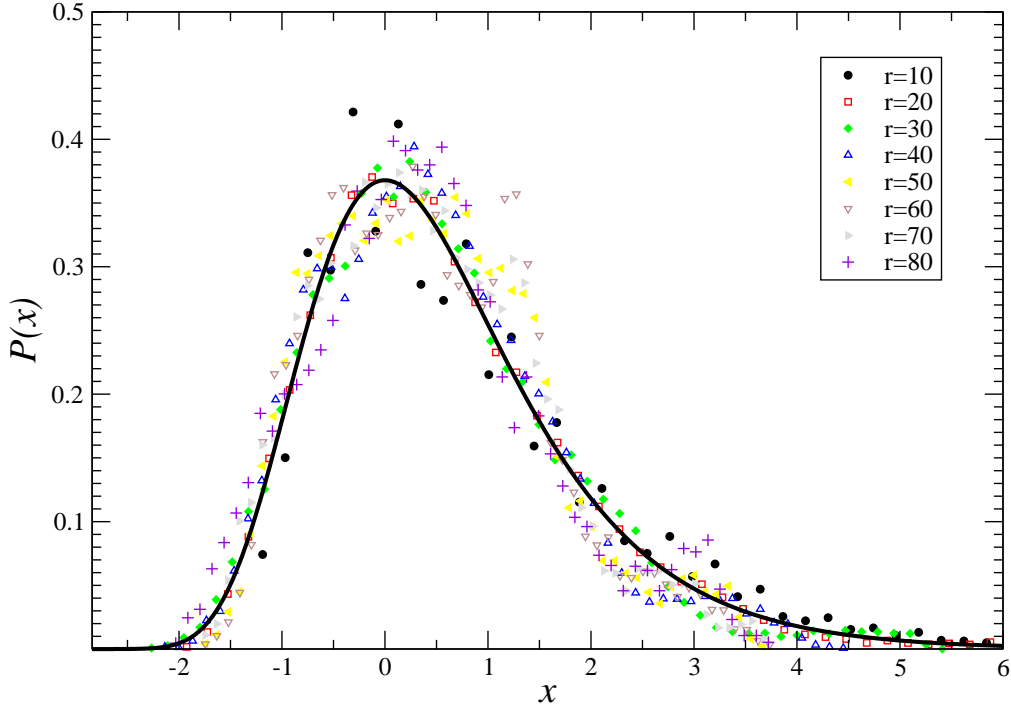}
\caption{
Data curves of different $r$ scaled together by fitting
  parameters $\alpha$ and $\beta$ for each curves. The solid line is
  the parameter-free Gumbel distribution Eq.\ref{gumbelscaled}.
(From \cite{gumbel}).}
\label{fig:scale}
\end{center}
\end{figure}

\begin{figure}
\begin{center}
\includegraphics*[angle=0, width=0.85\textwidth]{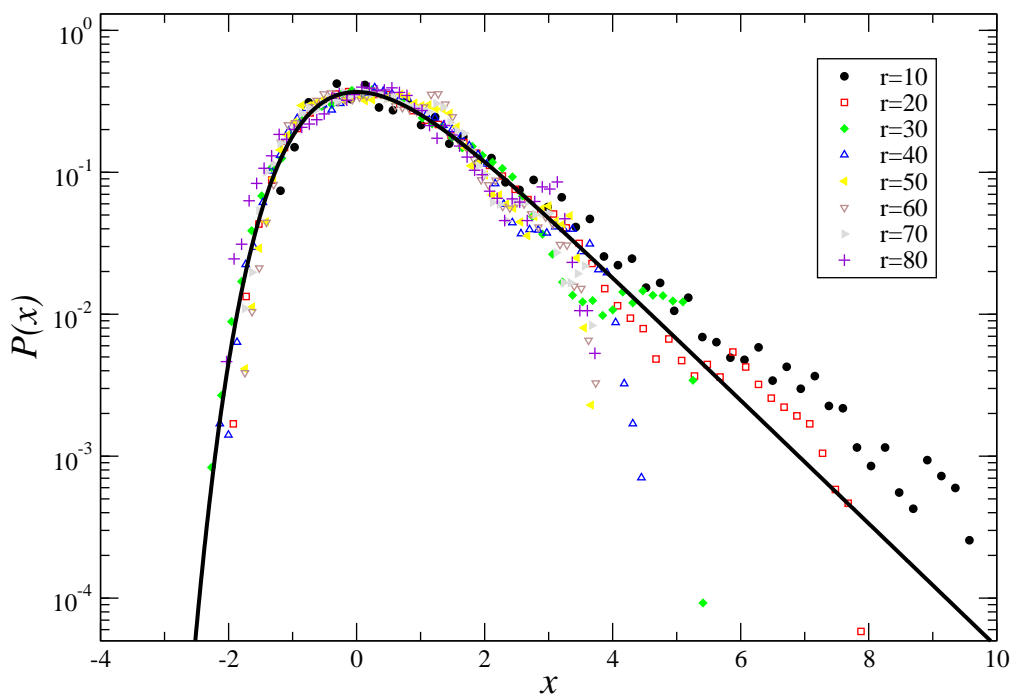}
\caption{The same as Fig.~\ref{fig:scale}, but on log-linear scale to
  emphasize the tails of the distribution.(From \cite{gumbel}).}
\label{fig:scalelog}
\end{center}
\end{figure}

To probe the whole distribution of the conditional density $n_i(r)$,
we fitted the Gumbel distribution (Eq.\ref{gumbel}) via its two
parameters $\alpha$ and $\beta$. One of our best fits is
  obtained for $r=20$ Mpc/h, see Fig.~\ref{fig:best20}. The data,
  moreover, convincingly collapses to the parameter-less Gumbel
  distribution (Eq.\ref{gumbelscaled}) for all values of $r$ for
  $10\le r\le 80$ Mpc/h, with the use of the scaling variable $x$ from
  Eq.\ref{scalevar} (see Figs.~\ref{fig:scale}-\ref{fig:scalelog}).
  Note that for a Poisson point process (uncorrelated random points)
  the number $N(r)$ (and consequently also the density) fluctuations
  are distributed exactly according to a Poisson distribution, which
  in turn converges to a Gaussian distribution for large average
  number of points $\overline{N(r)}$ per sphere. In our samples,
  $\overline{N(r)}$ is always larger than 20 galaxies, where the
  Poisson and the Gaussian PDFs differ less than the uncertainty in
  our data. Note also that due to the central limit theorem, all
  homogeneous point distributions (not only the Poisson process) lead
  to Gaussian fluctuations.  Hence the appearance of the Gumbel
  distribution is a clear sign of inhomogeneity and large scale
  structures in our samples.

The fitting parameters in Eq.\ref{gumbel} varied with the radius $r$
approximately as
\begin{equation}
\alpha \approx \frac{0.007}{r^{0.21}}, \quad \beta \approx
\frac{0.035}{r} \;.
\end{equation}
although a logarithmic fit $\alpha\approx 0.0115/\log r$ cannot be
excluded either.  With the fitted values of $\alpha$ and $\beta$ we
recover the (directly measured) average conditional density of
galaxies through Eq.~\ref{cumu}. On the other hand, we have a
discrepancy when comparing the directly measured $\sigma^2$ to that
obtained from the Gumbel fits through Eq.~\ref{cumu}. The reason for
this discrepancy is that the uncertainty in the tail of the PDF
$P(n,r)$ is amplified when we directly calculate the second moment.

Due to the scaling and data collapse we argued that the large scale
galaxy distribution shows similarities with critical systems
\cite{gumbel}.  Here the galaxy density around each galaxy is
analogous to a random variable describing a spatially averaged
quantity in a volume. The average conditional galaxy density depends
on the volume size ($\sim r^3$) only logarithmically
$\overline{n(r)}\sim 1/ \log r$ from Eq.\ref{densscale}. According to
the conjecture of Bramwell for critical systems \cite{bramwell09}, if
a spatially averaged quantity depends only weakly (say
logarithmically) on the system size, the distribution of this quantity
follows the Gumbel distribution. This is indeed what we see in the
galaxy data. Hence our two observations about the average density and
the density distribution are compatible with the behavior of critical
systems in statistical physics. We note that standard models of galaxy
formation predict homogeneous mass distribution beyond $\approx 10$
Mpc/h \cite{cdm_theo,sdss_epl,sdss_aea}.  To explain our findings
about non-Gaussian fluctuations up to much larger scales presents a
challenge for future theoretical galaxy formation models (see
\cite{cdm_theo,sdss_epl,sdss_aea,2df_epl,2df_aea,gumbel} for more
details).

\section{Assumptions and basic principles in cosmology} 
\label{assumptions}

As we noticed in the introduction, a widespread idea in cosmology is
that the so-called concordance model of the universe combines {\it
  two} fundamental assumptions.  The first is that the dynamics of
space-time is determined by Einstein's field equations. The second is
that the universe is homogeneous and isotropic. This hypothesis,
usually called the Cosmological Principle, is though to be a {\it
  generalization} of the Copernican Principle that ``the Earth is not
in a central, specially favored position'' \cite{bondi,pedro2}. The
FRW model is derived under these two assumptions and it describes the
geometry of the universe in terms of a single function, the scale
factor, which obeys to the Friedmann equation \cite{weinberg}.  There
is a subtlety in the relation between the Copernican Principle (all
observes are equivalent and there are no special points and
directions) and the Cosmological Principle (the universe is
homogeneous and isotropic).  Indeed, the fact that the universe looks
the same, at least in a statistical sense, in all directions and that
all observers are alike does not imply spatial homogeneity of matter
distribution. It is however this latter condition that allows us to
treat, above a certain scale, the density field as a smooth function,
a fundamental hypothesis used in the derivation of the FRW metric.
Thus there are distributions which satisfy the Copernican Principle
and which do not satisfy the Cosmological Principle \cite{book}.
These are statistically homogeneous and isotropic distributions which
are also spatially inhomogeneous \footnote{Mandelbrot \cite{man82}has
  introduced a modified version of the Cosmological Principle, named
  Conditional Cosmological Principle, which is the analogous of the
  Copernican Principle discussed in the text.}.  Therefore the
Cosmological Principle represents a specific case, holding for
spatially homogeneous distributions, of the Copernican Principle which
is, instead, much more general.  Statistical and spatial homogeneity
refer to two different properties of a given density field. The
problem of whether a fluctuations field is compatible with the
conditions of the absence of special points and direction can be
reformulated in terms of the properties of the probability density
functional (PDF) which generates the stochastic field.

Matter distribution in cosmology then is considered to be a
realization of a {\it stationary} stochastic point process. This is
enough to satisfy the Copernican Principle i.e., that there are no
special points or directions; however this does not imply spatial
homogeneity. Spatially homogeneous stationary stochastic processes
satisfy the special and stronger case of the Copernican Principle
described by Cosmological Principle.  Indeed, isotropy around each
point together with the hypothesis that the matter distribution is a
smooth function of position i.e., that this is analytical, implies
spatial homogeneity. (A formal proof can be found in
\cite{straumann74}.)  This is no longer the case for a non-analytic
structure (i.e., not smooth), for which the obstacle to applying the
FRW solutions has in fact solely to do with the lack of spatial
homogeneity \cite{pwa}.

The condition of {\em spatial homogeneity} ({\em uniformity}) is
satisfied if the ensemble average density of the field $\langle \rho
\rangle$ is strictly positive.  We discussed in the previous sections
tests to establish whether a point distribution in a given sample
(i.e. galaxies) is spatially uniform. The additional test provided by
the analysis of the PDF of conditional fluctuations in disjointed
sub-samples, i.e. Eq.\ref{selfavtest}, allows us to determine a
distribution is also statistically stationary.  In the case in which
systematic differences are found to be important one may then further
test whether this is due to the breaking of self-averaging properties
for the presence of large-scale fluctuations or whether this is due to
the lack of translational invariance. A statistical test able to
distinguish between these two cases, can also give an information
about the validity of the Copernican principle as we discuss in the
next section \footnote{Note that the breaking of the condition of
  translational invariance may also occur in the presence of a
  redshift-dependent selection effect. Thus the the violation of the
  Copernican principle due to the intrinsic lack of statistical
  translational invariance can be concluded only if all redshift
  dependent selection effects are taken into account.}.

\subsection{Testing the Copernican and cosmological principles} 
\label{examples}

\begin{figure}
\begin{center}
\includegraphics*[angle=0, width=0.85\textwidth]{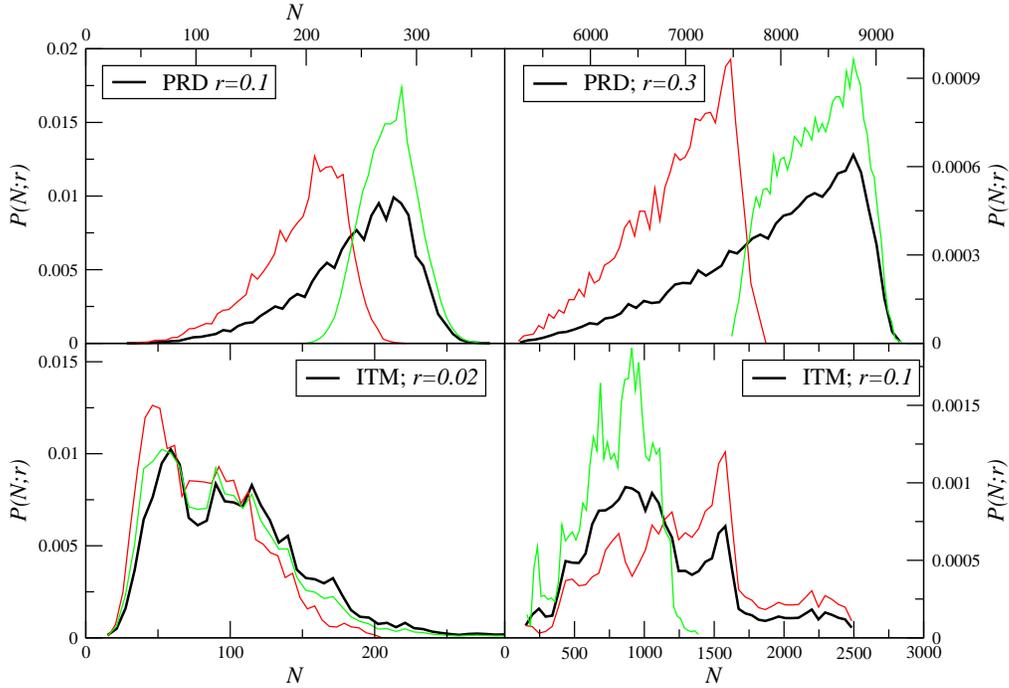} 
        \caption{Upper panels: The PDF for $r=0.1$ (left) and $r=0.3$
          (right) for PRD computed for the whole sample (black
          line). The red (green) line shows the PDF measured in the
          sub-sample placed closer to (father from) the origin. Lower
          Panels: The same for the ITM at scales $r=0.02$ (left) and
          $r=0.1$ (right) (From \cite{copernican}.}
	\label{pdf}
\end{center}
\end{figure}

Let us firstly consider a case where translational invariance is
broken. We generate a Poisson-Radial distribution (PRD) which is a
inhomogeneous distribution that can mimic the effect of a ``local
hole'' around the origin. In a sphere of radius $R_0=1$ we place, for
instance, $N=2\cdot 10^5$ points.  In each bin at radial distance from
the sphere center $[R_i,R_{i+1}]$, and with thickness $\Delta R$, the
distribution is Poissonian with a density varying as $n(R) = n_0 \cdot
R \;, $ where $n_0$ is a constant.
We determine the PDF $P(N;r)$ of conditional fluctuations obtained by
making an histogram of the values of $N(r;R)$ at fixed $r$ (see the
upper panels of Fig.\ref{pdf}).  The whole-sample PDF is clearly
left-skewed: this occurs because the peak of the PDF corresponds to
the most frequent counts which are at large radial distance simply
because shells far-way from the origin contain more points. The spread
of the PDF can easily be related to the difference in the density
between small and large radial distances in the sample.  By computing
the PDF into two non-overlapping sub-samples, nearby to and faraway
from the origin, one may clearly identify the systematic dependence of
this quantity on the specific region where this is measured. This
breaking of the self-averaging properties is caused by the
radial-distance dependence of the density and thus by the breaking of
translational invariance (as noticed above in the data a redshift
dependent selection effect may cause in the same result).

Let us now consider a stationary stochastic distribution, where the
breaking self-averaging properties is due to the effect of large scale
fluctuations. An example is represented by the inhomogeneous toy model
(ITM) constructed as follows. We generate a stochastic point
distribution by randomly placing, in a two-dimensional box of side
$L$, structures represented by rectangular sticks.  We first
distribute randomly $N_s$ points which are the sticks centers: they
are characterized by a mean distance $\Lambda \approx
(L^3/N_s)^{1/3}$.  Then the orientation of each stick is chosen
randomly. The points belonging to each stick are also placed randomly
within the stick area, that for simplicity we take to be $\ell \times
\ell/10$.  The length-scale $\ell$ can vary, for example being
extracted from a given PDF. The number of sticks placed in the box
fixes $\Lambda$.  This distribution is by construction stationary
i.e., there are no special points or directions.  When $\ell \ge L$
and $\Lambda \le L$ but with $\ell$ varying in such a way that there
can been large differences in its size, the resulting distribution is
long-range correlated, spatially inhomogeneous and it can be not
self-averaging.  This latter case occurs when, by measuring the PDF of
conditional fluctuations in different regions of a given sample, one
finds, for large enough $r$, systematic differences in the PDF shape
and peak location (see the bottom panels Fig.\ref{pdf}). These are due
to the strong correlations extending well over the size of the sample.

How can we distinguish between the case in which a distribution is not
self-averaging because it is not statistically translational invariant
and when instead this is stationary but fluctuations are too extended
in space and have too large amplitude ?  The clearest test is to
change the scale $r$ where $P(N,r)$ is measured, and determining
whether the PDF is self-averaging. Indeed, in the case of the PRD the
strongest differences between the PDF measured in regions placed at
small and large radial distance from the structure center, occur for
small $r$. This is because the local density has the largest
variations at small and large radial distances by construction.  When
$r$ grows, different radial scales are mixed as the generic sphere of
radius $r$ pick up contributions both from points nearby the origin
and from those far away from it, resulting in a smoothing of local
differences.  Instead, in the ITM for small $r$ the difference is
negligible while for large enough $r$ the different determinations of
the local density start to feel the presence of a few large structures
which dominate the large scale distribution in the sample.

\subsection{Implications for theoretical modeling} 

The discussion in the previous sections was meant to treat the
statistical properties of the galaxy density field in a spatial
hyper-surface.  As mentioned above, this is an approximation valid
when considering the galaxy distribution limited to relatively low
redshifts, i.e. $z<0.2$. In particular, we have developed a test to
focus on the properties of statistical and homogeneity homogeneity in
nearby redshift surveys. The assumptions of the cosmological model
enter in the data analysis when calculating the metric distance from
the redshift and the absolute magnitude from the apparent one and the
redshift. However, given that second order corrections are small for
$z<0.2$, our results are basically independent on the chosen
underlying model to reconstruct metric distances and absolute
magnitudes from direct observables. In practice we can use just a
linear dependence of the metric distance on the redshift (which is, to
a very good approximation, compatible with observations at low
redshift). For this same reason we can approximate the observed
galaxies as lying in a spatial hyper-surface.

In the ideal case of having a very deep survey, up to $z\approx 1$, we
should consider that we make observations on our past light-cone which
is not a space-like surface. In order to evolve our observations onto
a spatial surface we would need a cosmological model, which at such
high redshift can play an important role in the whole determination of
statistical quantities.  A sensible question is whether we can to
reformulate the statistical test given so that it can be applied to
data on our past light-cone, and not on an assumed spatial
hyper-surface. Going to higher redshift poses a number of question,
first of the all the one of checking the effect of the assumptions
used to construct metric distances and absolute magnitudes from direct
observables. Testing these effects can be simply achieved by using 
different distance-redshift relations. 

However, we note that a smooth change of the distance-redshift
relation as implied by a given cosmological model, may change the
average behavior of the conditional density as a function of redshift
but it cannot smooth out fluctuations, i.e. it cannot substantially
change the PDF of conditional fluctuations when they are measured
locally. Indeed, our test is based on the characterization of the PDF
of conditional fluctuations and not only of the behavior of the
conditional average density as a function of distance. The PDF
provides, in principle, with a complete characterization of the
fluctuations statistical properties.  We have shown that the PDF of
fluctuations has a clear imprint when the distribution is spherically
symmetric or when it is spatially inhomogeneous but statistically
homogeneous.

 The fact that we analyze conditional fluctuations means that we
 consider only local properties of the fluctuations: local with
 respect to an observer placed at different radial (metric) distances
 from the us, i.e. at different redshifts. For the determination of
 the PDF we have to consider two different length scale: the first is
 the (average) metric distance $R$ of the galaxies on which we center
 the sphere and the second is the sphere radius $r$. Irrespective of
 the value of $R$ when $r$ is smaller than a few hundreds Mpc (i.e.,
 when its size is much smaller than any cosmological length scale), we
 can always locally neglect the specific $R(z)$ relation induced by a
 specific cosmology.  In other words, when the sphere radius is
 limited to a few hundreds Mpc we can approximate the measurements of
 the conditional density to be performed on a spatial hyper-surface.

The whole description of the matter density field in terms of FRW or
even Lemaitre-Tolman-Bondi (LTB) cosmologies, refer to the behavior
of, for instance, the average matter density as a function of time (in
the LTB case also as a function of scale) but it says anything on the
fluctuation properties of the density field.  Thus, when looking at
different epochs in the evolution of the universe, we should detect
that the average density varies (being higher in early epochs). This
means only that the peak of the PDF will be located at different $N$
values, but the shape of the PDF is unchanged by this overall (smooth)
evolution. Fluctuations are simply not present in the FRW or LTB
models, and the whole issue of back-reaction studies is to understand
what is their effect.

Note that models which explain dark energy through inhomogeneity do so
using a spatial under-density in the matter density which varies on
Gpc scales --- out to $z \approx 1$ \cite{pedro}. These models by
placing us at the center of the universe, violate the Copernican
Principle. In this respect we note that, while we cannot make any
claim for $z>0.2$ based on current data, the fact that galaxy
distribution is spatially inhomogeneous but statistically homogeneous
up to 100 Mpc/h, already poses intriguing theoretical
problems. Indeed, in that in that range of scales, the modeling of the
matter density field as a perfect fluid, as required by the FRW
models, is not even a rough approximation. As pointed out by various
authors \cite{slmp98,wiltshire3}, if the linearity of the Hubble law
is a consequence of spatial homogeneity, how is it that observations
show that it is very well linear at the same scales where matter
distribution is inhomogeneous ?  Recently \cite{wiltshire2} it was
speculated a solution to this apparent paradox can be found by
considering both the effects of back-reaction and the synchronization
of clocks. While this is certainly an interesting approach, the
formulation of a more complete and detailed theoretical framework is
still lacking.

Finally we note that there are several complications in radially
 inhomogeneous models at high redshift. Beyond the change of the
 distance-redshift relation, discussed above, another is how structure
 evolves from our past light-cone onto a surface of constant time. Thus
 in order to make a precise test on the spatial properties of a given
 model, one needs to develop the corresponding theory of structure
 formation. However, at least at low redshifts, it seems implausible
 that the main feature of the model, the specific redshift-dependence
 of the spatial density, will not be the clearer prediction for 
the observations of galaxy structures.

\section{Conclusions} 
\label{conclusions}  

We discussed several results showing that galaxy distribution in the
newest galaxy samples is characterized by large fluctuations. These
are manifest in the scaling properties of the conditional density
which shows scaling behaviors. Particularly at small scales this
statistics presents a power-law behavior with an exponent close to
minus one, corresponding to a fractal dimension $D\approx 2$.  The
difference with the different dimension $D=1.2$ reported by authors
(see e.g. \cite{dp83,davis88,park}) is due to the finite size effects
which perturb the estimation of the two-point correlation function
$\xi(r)$ \cite{book}.

On larger scales and up to $r\approx 80$ Mpc/h a smaller correlation
exponent is found to fit the data: the density depends, for $20 \le
r\le 80$ Mpc/h, only weakly (logarithmically) on the system
size. Correspondingly, we find that the density fluctuations follow
the Gumbel distribution of extreme value statistics. This distribution
is clearly distinguishable from a Gaussian distribution, which would
arise for a homogeneous spatial galaxy configuration. While in the
Gaussian case the rapid decay of the tails of the distribution cut
large fluctuations, in the Gumbel case the large value tail decays
slower. In such a situation the density field is still inhomogeneous,
although not as wild as in the case in which the PDF presents
power-law tails.

We discussed that in several samples it is found that self-averaging
properties, at large scales, are not satisfied.  This is due to the
presence of large scale galaxy structures which correspond to density
fluctuations of large amplitude and large spatial extension, whose
size is limited only by the sample boundaries.  Note that the lack of
self-averaging does not allow one to characterize the nature of
fluctuations; this is however a clear indication that the distribution
has not reached spatial uniformity.

The large scale inhomogeneities detected in the three dimensional
galaxy samples are at odds with the predictions of standard models. In
particular according to these models the density field should present
on large scales sub-Poissonian fluctuations, or a super-homogeneous
nature with negative correlations \cite{glass,book}.  Forthcoming
redshift surveys will allow us to clarify whether on such large scales
galaxy distribution is still inhomogeneous but statistically
stationary, or whether the evidences for the breaking of spatial
translational invariance found in the galaxy samples considered were
due to selection effects in the data.

Finally we discussed that the galaxy distribution is found to be
compatible with the assumptions that this is transitionally invariant,
i.e. it satisfies the requirement of the Copernican Principle that
there are no spacial points or directions., while because of lack of
spatial homogeneity galaxy distribution is not compatible with the
stronger assumption of spatial homogeneity, encoded in the
Cosmological Principle.

\subsection*{Acknowledgments}
We are in debt with T. Antal, Y. Baryshev and
N. L. Vasilyev for fruitful collaborations. We also thank A.
Gabrielli and M. Joyce for interesting discussions.  We acknowledge
the use of the Sloan Digital Sky Survey data ({\tt
  http://www.sdss.org}).

\end{document}